\documentclass[aip,cha,amsmath,amssymb,reprint,author-numerical]{revtex4-1}

\usepackage{graphicx}
\usepackage{dcolumn}
\usepackage{bm}
\usepackage{hyperref}
\usepackage{color}
\begin{document}


\title{Control of birhythmicity: A self-feedback approach} 



\author{Debabrata Biswas}
\email{debbisrs@gmail.com}
\affiliation{Chaos and Complex Systems Research Laboratory, Department of Physics, University of Burdwan, Burdwan 713 104, West Bengal, India}
\author{Tanmoy Banerjee}
\email{tbanerjee@phys.buruniv.ac.in}
\thanks{Corresponding author}
\affiliation{Chaos and Complex Systems Research Laboratory, Department of Physics, University of Burdwan, Burdwan 713 104, West Bengal, India}
\author{J{\"u}rgen Kurths}
\email{juergen.kurths@pik-potsdam.de}
\affiliation{Potsdam Institute for Climate Impact Research, Telegraphenberg, D-14415 Potsdam, Germany}
\affiliation{Institute of Physics, Humboldt University Berlin, D-12489 Berlin, Germany}
\affiliation{Institute for Complex Systems and Mathematical Biology, University of Aberdeen, Aberdeen AB24 3FX, United Kingdom}
\affiliation{Institute of Applied Physics, Russian Academy of Sciences, 603950 Nizhny Novgorod, Russia}


\received{:to be included by reviewer}
\date{\today}

\begin{abstract}
Birhythmicity occurs in many natural and artificial systems. In this paper we propose a self-feedback scheme to control birhythmicity. To establish the efficacy and generality of the proposed control scheme, we apply it on three birhythmic oscillators from diverse fields of natural science, namely, an energy harvesting system, the p53-Mdm2 network for protein genesis (the OAK model) and a glycolysis model (modified Decroly-Goldbeter model). Using the harmonic decomposition technique and energy balance method we derive the analytical conditions for the control of birhythmicity. A detailed numerical bifurcation analysis in the parameter space establishes that the control scheme is capable of eliminating birhythmicity and it can also induce transitions between different forms of bistability. As the proposed control scheme is quite general, it can be applied for control of several real systems, particularly in biochemical and engineering systems.
\end{abstract}


\maketitle 

\begin{quotation}
Multistability appears in diverse forms and their study is an exciting topic of research in science and engineering. A particular form of multistability is bistability: it shows many variant, such as, the coexistance of two stable steady states, one stable steady state and one stable limit cycle, two stable limit cycles, or two chaotic attractors. Birhythmicity is the phenomenon of coexistence of two stable limit cycles separated by an unstable limit cycle with different amplitudes and frequencies. In many physical systems birhythmicity is undesirable as in energy harvesting systems but in most biological systems, e.g., enzymatic oscillations, it is desirable. Therefore, control of birhythmicity is of utmost importance. Although the control of multistability is a well studied topic, the control of birhythmicity has not been explored to that extent. In this paper we propose a control scheme that can effectively control and, whenever required, can eliminate birhythmicity. We theoretically explore and numerically establish the technique of control of birhythmicity and transitions to any desired attractor. A number of engineering and biological systems are investigated with the proposed control scheme to establish the efficacy and generality of the scheme. The main essence of this control scheme lies in the fact that it is easily realizable and offers an efficient mean to control birhythmicity.
\end{quotation}

\section{Introduction}
\label{sec:intro}
Birhythmicity is a variant of multistability \cite{Pisarchik14}, which arises in many natural and artificial systems in the field of physics\cite{Kwuimy15}, biology\cite{Goldbeter96, Doldbeter02nat, Dsray04} and chemistry\cite{Alamgir83}. The coexistence of two stable limit cycles of different amplitudes (and frequencies) separated by an unstable limit cycle is the signature of birhythmicity. Birhythmicity may appear in chaotic oscillators also, e.g., the coexistence of two chaotic attractors has been reported in the literature\cite{chua_co} and studied in detail in Ref.~\onlinecite{Pisarchik06}. The appearance of birhythmicity plays a crucial role in living systems as it helps to maintain different modes of oscillations that organize various biochemical processes in response to variations in their environment \cite{Alamgir83}. That is why most of the biochemical oscillators are birhythmic. A few prominent examples are: glycolytic oscillator and enzymatic reactions \cite{Goldbeter96,Doldbeter02nat,karroy03_prl,Dsray04}, intracellular $\mbox{Ca}^{2+}$ oscillations \cite{Pisarchik06}, birhythmic oscillations due to the complex regulatory properties of allosteric enzymes, namely phosphofructokinase (PFK), which is activated by ADP and feedback due to this ADP to ATP\cite{Dsray04}, birhytmicity in the p53-Mdm2 network\cite{Abou09, Vogelstein00}, oscillatory generation of cyclic AMP during the aggregation of slime mold {\em Dictyostelium discodeum}\cite{Martiel87} or circadian oscillation in PER and TIM proteins in Drosophila\cite{Leloup99}. Unlike in living systems, birhythmicity is often undesirable in physical \cite{jj14} and engineering systems \cite{Kwuimy15}. For example, birhythmicity limits the efficiency of an energy harvesting system \cite{Kwuimy15}: In an energy harvesting system the wind-induced vibration shows birhythmic oscillations, thus, depending upon the initial vibrational energy of the wind the system may oscillate in  a small amplitude limit cycle and results in less mechanical deformation that, in turn, yields less electrical energy. Therefore, for an energy harvesting system the oscillation with a large amplitude limit cycle is always desirable to have larger energy production. Clearly in some situations birhythmicity is undesirable while in others it is a necessity. This marks the importance of control of birhythmicity.

A recent extensive review work on control of multistability by \citet{Pisarchik14} suggests that although several control mechanisms have been reported for the control of bistable systems containing oscillations and stable steady state\cite{Pisarchik00,Pisarchik01}, the control of birhythmicity is a less explored topic. Only a few works are reported on the control of bithythmicity. \citet{Ghosh11} showed  that time delay feedback control, which was originally proposed by \citet{Pyragas95} to control chaos, is able to control birhythmicity as well in a modified birhythmic van der Pol oscillator. However, owing to the presence of time delay in their control scheme a detailed bifurcation analysis for the controlled system is a difficult task. Also, the implementation of a delayed signal to be fed is challenging. \citet{Sevilla15} showed that application of a harmonic modulation and the presence of a positive feedback along with a proper choice of the parameters can transform a multistable system with coexisting periodic and chaotic attractors to a monostable one. Recently, in Ref.~\onlinecite{Biswas2016control} we have proposed a technique to control birhythmicity by using a conjugate self-feedback method. This technique was verified using a variant of the van der Pol oscillator with birhythmic oscillations and it was shown that the conjugate self-feedback in that oscillator is capable of removing birhtymicity by inducing monorhythmic oscillation. However, in our scheme in Ref.~\onlinecite{Biswas2016control} one requires the access of two variables: the variable of interest and its conjugate counterpart.

In this paper we propose a more general and experimentally feasible control technique that employs only one accessible variable. We establish the effectiveness of this control technique using three real systems from diverse field of physics and biology, namely (i) an energy harvesting system \cite{Kwuimy15}, (ii) the p53-Mdm2 network popularly known as the OAK model \cite{Abou09} and (iii) a variant of glycolytic oscillators \cite{Goldbeter96,Dsray04}. The control and taming of birhythmic oscillations in these three oscillators of different origin also establish the generality of our control scheme. To establish the efficacy of our scheme, we carry out an extensive theoretical analysis using the harmonic decomposition technique and the energy balance method. Also, we employ a rigorous numerical bifurcation analysis to identify the parametric zone of occurrence of bi- and mono-rhythmic oscillations and their exact genesis.

The paper is organized in the following manner: The next section describes the details of control of birhythmicity in an  energy harvesting model. We carry out a detailed analysis for the onset of birhythmicity. In sections \ref{sub:oak} and \ref{sub:glyco} we consider the control of birhythmicity in the p53-Mdm2 network (OAK model) and glycolytic oscillator, respectively.  Finally, section~\ref{sec:conc} concludes the outcomes of the study.


\section{Energy harvesting system}
\label{sub:energy}
\subsection{The model}
\begin{figure}
\includegraphics[width=0.49\textwidth]{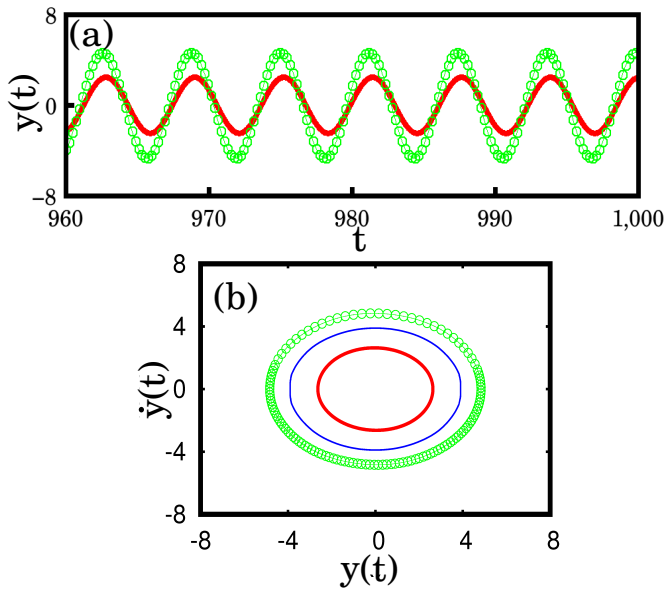}
\caption{(a) Time series and (b) phase plane plot of the energy harvesting system \eqref{energyeq0}. The small amplitude LC is for ICs $\mathcal{I}_1$ and large amplitude LC is that for $\mathcal{I}_2$ (see text). The trajectory in blue in the middle represents the unstable LC.  Other parameters are: $\mu=0.1$, $\alpha=0.144$, $\beta=0.005$, $\Omega_0=1$, $\eta_0=0.1$, $\eta_1=0.25$ and $\gamma=0.2$}
\label{energy_bi_ts}
\end{figure}

At first we describe the original model of an energy harvesting system. Energy harvesting systems generate electrical energy from ambient energy arising from sources like structural vibration, wind flow, physiological and chemical reactions, etc. \cite{Abdelkefi13,*Kwon10,*Kwuimy14pla,*Kwuimy12,*Litak12,*Tekam14,*Tekam15}. \citet{Kwuimy15} considered an energy harvesting model which is implemented with an arrangement of a cantilever attached to piezoelectric patches under the action of transverse wind flow. The physical model consists of an electrical circuit with a load resistance and a flexible beam of distributed piezoelectric patches. The dimensionless form of the original model is given by the following set of equations:
\begin{subequations}\label{energyeq0}
\begin{align}
\frac{d^2 y}{dt^2}+\mu F\bigg(\frac{dy}{dt}\bigg)+\Omega_0^2 y&=\eta_0 v,\label{energyeqa0}\\
\frac{dv}{dt}+\gamma v&=-\eta_1\frac{dy}{dt},\label{energyeqb0}
\end{align}
\end{subequations}
Here $y$ is the dimensionless transversal beam deflection function and $v$ is the dimensionless form of the voltage generated by the piezoelectric element. Also, $\mu$, $\Omega_0$, $\gamma$, $\eta_0$ and $\eta_1$ are all positive parameters and the nonlinear function $F\bigg(\frac{dy}{dt}\bigg)$ is given by
\begin{equation}\label{energyfn}
F\bigg(\frac{dy}{dt}\bigg)=-\frac{dy}{dt}+\frac{1}{3}\bigg(\frac{dy}{dt}\bigg)^3-\frac{\alpha}{5}\bigg(\frac{dy}{dt}\bigg)^5+\frac{\beta}{7}\bigg(\frac{dy}{dt}\bigg)^7.
\end{equation}
The system is birhythmic for the following parameters\cite{Kwuimy15}: $\mu=0.1$, $\alpha=0.144$, $\beta=0.005$, $\Omega_0=1$, $\eta_0=0.1$, $\eta_1=0.25$ and $\gamma=0.2$. It shows three distinct limit cycles (LCs) (two stable and one unstable) depending upon two sets of initial conditions (IC), namely, $\mathcal{I}_1\equiv \big(y(0),dy(0)/dt,v(0)\big)\in (0.1, 0, 0.3)$ (small amplitude stable LC) and $\mathcal{I}_2\equiv \big(y(0),dy(0)/dt,v(0)\big)\in (7, 0, 0.3)$ (large amplitude stable LC). The unstable LC determines the basin boundary of these stable LCs. The time series and phase plane plots for these sets of initial conditions are shown in Fig.~\ref{energy_bi_ts}(a) and Fig.~\ref{energy_bi_ts}(b), respectively. From Fig.~\ref{energy_bi_ts}(b) we see that the system shows two limit cycles separated by an unstable LC indicating birhythmicity. 
{\color{black}
\subsection{The self-feedback control scheme}
We consider a dynamical system modeled by the following equation:
\begin{equation}\label{sf}
\ddot{x}+\mu F(\dot{x})+x=G(x,\dot{x},t).
\end{equation}
Where $x\in \mathbb{R}$, $F(\dot{x})$ is a nonlinear function, and $\mu$ is a system parameter that determines the intrinsic dynamics of the considered system. Here $G(x,\dot{x},t)$ is the control term. In our present case it is given by $G(x,\dot{x},t)=-d\dot{x}$, where $d$ controls the strength of the self-feedback term. Note that to implement the system one needs the access to a single (scalar) variable. This type of self-feedback has been used and implemented earlier in the Fabry-Perot laser diode system \cite{self_laser}, biological system to control complex motor task \cite{self_bio1} and learning \cite{self_bio2}. Another variant of the self-feedback control is the paradigmatic Pyragas control technique \cite{Pyragas95} where an additional time-delayed version of the variable is used. In this paper our goal is to study the effect of self-feedback term on birhythmicity; we concentrate on the fact that how the self-feedback  parameter $d$ affects the birhythmic oscillation. Since birhythmicity involves global bifurcations therefore we use continuation based rigorous bifurcation analysis along with theoretical analysis to track the complete system behavior.} 
\subsection{Control of the energy harvesting system with self-feedback}
\label{subsub:dnzero}
Next we apply the self-feedback control scheme to \eqref{energyeq0}, which now reads
\begin{subequations}\label{energyeq}
\begin{align}
\frac{d^2 y}{dt^2}+\mu F\bigg(\frac{dy}{dt}\bigg)+\Omega_0^2 y&=\eta_0 v-d\frac{dy}{dt},\label{energyeqa}\\
\frac{dv}{dt}+\gamma v&=-\eta_1\frac{dy}{dt},\label{energyeqb}
\end{align}
\end{subequations}
The term $-d\frac{dy}{dt}$ in \eqref{energyeqa} represents the self-feedback proportional to the time rate of change of the transversal beam deflection with $d$ as the self-feedback strength. $d$ also determines the nature of the self-feedback: $d>0$ represents a positive feedback, and $d<0$ represents a negative one and $d=0$ implies no feedback. 

In order to analyze the controlled system, we reduce Eq.~\eqref{energyeq} to the following single equation
\begin{equation}\label{eneqd}
\ddot{y}+\mu F(\dot{y})+\Omega_0^2 y+d\dot{y}+\frac{\eta_0}{\gamma}(\eta_1 \dot{y}+\dot{v})=0.
\end{equation}
According to Refs.~\onlinecite{Kwuimy15, Kwuimy14}, the variable $v$ does not affect the dynamical property of the fixed point of the system and hence they showed that one can put $\eta_0=\eta_1=0$. Then Eq.~\eqref{eneqd} becomes
\begin{equation}\label{energycoupsingle}
\ddot{y}+\mu F(\dot{y})+\Omega_0^2 y+d\dot{y}=0.
\end{equation}
{\color{black} Next, we will constitute the {\it amplitude equation} of Eq.~\ref{energycoupsingle} to predict the kind of bifurcation structures associated with the system. Generally, for this, the most common technique in the literature is to apply the harmonic decomposition technique (see for example Refs.\onlinecite{Yamapi07,Ghosh11,Biswas2016control}). Although the harmonic decomposition technique is not an asymptotic method, however it can predict the amplitude equation in a simple yet effective way. In this context one more technique, which is a much more suitable technique for weakly nonlinear systems, is the Poincar\'e-Lindstedt technique which is discussed in Appendix-\ref{app:plm}; it is shown that both the analyses give equivalent {\it amplitude equations} and match well with the numerical results.}

According to the harmonic decomposition technique we assume the approximate solution of Eq.~\eqref{energycoupsingle} as
\begin{equation}\label{ysol}
y(t)=A\cos(\omega t),
\end{equation}
where $A$ is the amplitude and $\omega$ is the frequency of the oscillator with feedback. Substitution of this in Eq.~\eqref{energycoupsingle} yields
\begin{equation}\label{energyampfull}
\begin{split}
(&-\omega^2+\Omega_0^2)A\cos(\omega t)\\
=&\mu\omega\bigg(-1+\frac{1}{4}\omega^2 A^2-\frac{\alpha}{8}\omega^4 A^4+\frac{5\beta}{64}\omega^6 A^6\bigg)A\sin(\omega t)\\
&+d\omega A\sin(\omega t)\\
&+\mu\omega^3\bigg(-\frac{1}{12}+\frac{\alpha}{16}\omega^2 A^2-\frac{3\beta}{64}\omega^4 A^4\bigg)A^3\sin(3\omega t)\\
&+\mu\omega^5\bigg(-\frac{\alpha}{80}+\frac{\beta}{64}\omega^2 A^2\bigg)A^5\sin(5\omega t)\\
&-\frac{\beta}{448}\mu\omega^7 A^7 \sin(7\omega t).
\end{split}
\end{equation}

According to Ref.~\onlinecite{Jordan99} we can treat the higher harmonic terms as forcing terms and ignore them. Thus Eq.~\eqref{energyampfull} reduces to
\begin{equation}\label{energyhar}
\begin{split}
(&-\omega^2+\Omega_0^2)A\cos(\omega t)\\
=&\mu\omega\bigg(-1+\frac{1}{4}\omega^2 A^2-\frac{\alpha}{8}\omega^4 A^4+\frac{5\beta}{64}\omega^6 A^6\bigg)A\sin(\omega t)\\
&+d\omega A\sin(\omega t)+\mathcal{H},
\end{split}
\end{equation}
where $\mathcal{H}$ denotes higher harmonic terms.

From Eq.~\eqref{energyhar} we get the following frequency and amplitude equations:
\begin{equation}\label{freqeq}
\omega^2-\Omega_0^2=0,
\end{equation}
and
\begin{equation}\label{ampeq}
\mu\bigg(1-\frac{1}{4}\omega^2 A^2+\frac{\alpha}{8}\omega^4 A^4-\frac{5\beta}{64}\omega^6 A^6\bigg)-d=0.
\end{equation}
From Eq.~\eqref{ampeq} we infer that it is not a purely amplitude equation as it contains the frequency $\omega$ in it. So we substitute the value of $\omega$ from Eq.~\eqref{freqeq} in \eqref{ampeq} and get
\begin{equation}\label{ampeq1}
\mu\bigg(1-\frac{1}{4}\Omega_0^2 A^2+\frac{\alpha}{8}\Omega_0^4 A^4-\frac{5\beta}{64}\Omega_0^6 A^6\bigg)-d=0.
\end{equation}
It may be noted that for $d=0$ Eq.~\eqref{ampeq1} represents the amplitude equation of the uncontrolled energy harvesting system. It is interesting to note that the amplitude of the system does not depend on $\mu$ unless $d\neq 0$. The frequency in the harmonic limit becomes $\omega=1$. Also the frequency equation viz. Eq.~\eqref{freqeq} states that the frequency of the system does not depend on the feedback strength $d$ (unlike Ref.~\onlinecite{Biswas2016control}), thus, leaving the original frequency of the system intact. The three roots of Eq.~\eqref{ampeq1} corresponds to the amplitudes of the three limit cycles (two stable and one unstable). 

To test the stability of the system, we apply the energy balance method as suggested in Ref~\onlinecite{Ghosh11}. For $\mu=0$ and $d=0$, the harmonic solution of Eq.~\eqref{energycoupsingle} may be given as \cite{Yamapi07}
\begin{equation}
y(t)=A\cos(t+\phi),
\end{equation}
where $\phi$ is the initial phase and may be considered $\phi=0$ for convenience. The phase plane of this is a circle with period $T=2\pi$. In the presence of a self-feedback we can approximate
\begin{equation}\label{ycos}
y(t)\backsimeq A\cos(t),
\end{equation}
Let us consider $\big(-\mu F(\dot{y})-d(\dot{y})\big)$ as the external forcing term to calculate the change in energy $\Delta E$ during one period, i.e., $0\leq t\leq T$, with $T=2\pi$. The change in energy is given by
\begin{eqnarray}\label{inteq}
\Delta E&=&E(T)-E(0),\nonumber\\
&=&\int_0^T \big(-\mu F(\dot{y})-d(\dot{y})\big)\dot{y}dt.
\end{eqnarray}
$\Delta E=0$ for a periodic solution (limit cycle). Hence from the above integral we get using the condition of Eq.~\eqref{ycos}
\begin{equation}\label{fasq}
f(A^2)=\mu\bigg(1-\frac{1}{4}A^2+\frac{\alpha}{8}A^4-\frac{5\beta}{64}A^6\bigg)-d=0.
\end{equation}
Note that, Eq.~\eqref{fasq} is equivalent to Eq.~\eqref{ampeq} for $\omega=1$. The number of limit cycles can be obtained by solving Eq.~\eqref{fasq} by normalizing the frequency to unity. The number of positive roots gives the number of LCs. The stability of the limit cycle is determined by the slope of the curve of Eq.~\eqref{fasq} at the zero crossing point. The negativity of the slope determines the stability of the LC. The condition of stable limit cycle thus can be written as
\begin{equation}\label{deda}
\frac{d}{dA}\bigg(\Delta E(A)\bigg)\bigg\rvert_{\mbox{limit cycle}}< 0.
\end{equation}
\begin{figure}
\includegraphics[width=0.49\textwidth]{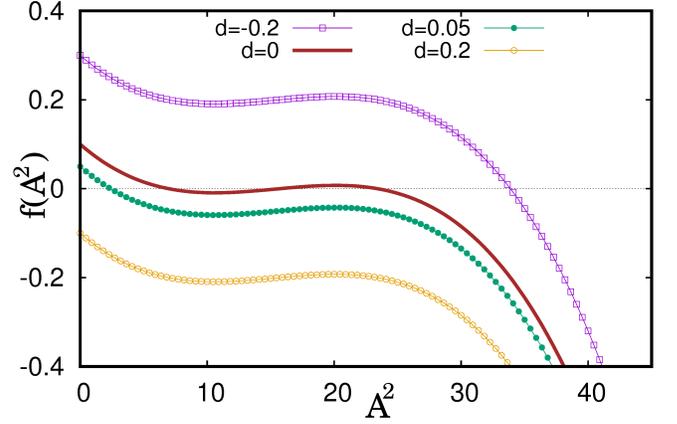}
\caption{Plot of $f(A^2)-A^2$ for the parameters $\mu=0.1$, $\alpha=0.144$, $\beta=0.005$, $\Omega_0=1$ for different values of the control parameter $d$. The line with hollow squares represents the case of a single LC with large amplitude for $d=-0.2$. The solid brown line shows the case of birhythmic oscillations for $d=0$. The line with solid circles gives the monorhythmic case with small amplitude for $d=0.05$ and the line with hollow circles (lower one) represents the case of stable steady states for $d=0.2$.}
\label{energy_ava}
\end{figure}
We solve the amplitude equation \eqref{fasq} (which is the same as Eq.~\eqref{ampeq1} for $\Omega_0=1$) by graphical method, i.e., we plot the polynomial $f(A^2)$ with $A^2$. The zero crossing points of $f(A^2)$ are the solutions of the equation. Here we consider the following parameter values: $\mu=0.1$, $\alpha=0.144$, $\beta=0.005$. These are the values of the parameters for which the original system \eqref{energyeq} exhibits birhythmicity in the absence of self-feedback. Now we vary the control parameter $d$ to get different solutions. The number of solutions of Eq.~\eqref{fasq} determines the number of limit cycles. The solutions for different values of the control parameter $d$ is shown in Fig.~\ref{energy_ava}. We vary $d$ from positive high to negative high (curves in Fig.~\ref{energy_ava} from lower o upper). From the figure we see that there is no zero crossing for $d=0.2$ (yellow line with hollow circles). Thus there is no limit cycle for this value of the control parameter. In other words at $d=0.2$ the system is in a stable steady state (SSS). Decrease of $d$ causes $f(A^2)$ to shift upwards and eventually to cross the zero line, thus giving rise to a stable limit cycle. This case is shown for $d=0.05$ in the figure (green line with solid circles). At this value of the parameter there is only one stable limit cycle with low amplitude. The negative slope of the curve at the zero crossing indicates that the the LC is stable. Next, at $d=0$ (solid line) there are three zero crossing points. Two of them have negative slopes and one has positive slope at the zero crossing points. Thus, there exists birhythmicity with two stable LCs (with negative slopes) and one unstable LC (with positive slopes) in between them. At $d=-0.2$ (purple line with hollow squares) there is only one zero crossing of $f(A^2)$ with a negative slope at the zero crossing exhibiting the presence of only one stable LC with larger amplitude. 

In the absence of control the system undergoes only a global bifurcation, namely, a saddle-node bifurcation of limit cycle (SNLC) and a codimension-2 cusp bifurcation in the $\alpha-\beta$ parameter space. However, the presence of the control term causes the system to experience a local bifurcation, namely, a Hopf bifurcation (HB). The eigenvalues of the Jacobian of Eq.~\eqref{energycoupsingle} around the stable points $(y,\dot{y})=(0,0)$ are given by 
\begin{equation}\label{ev}
\lambda_{1,2}=\frac{1}{2}\bigg[(\mu-d)\pm\sqrt{(d-\mu)^2-4\Omega_0}\bigg],
\end{equation}
and the condition of Hopf bifurcation reads
\begin{equation}\label{hopfeq}
d_{\mbox{HB}}=\mu,
\end{equation}
where $d_{\mbox{HB}}$ is the critical value of $d$ for a Hopf bifurcation to occur.

\begin{figure}
\includegraphics[width=0.49\textwidth]{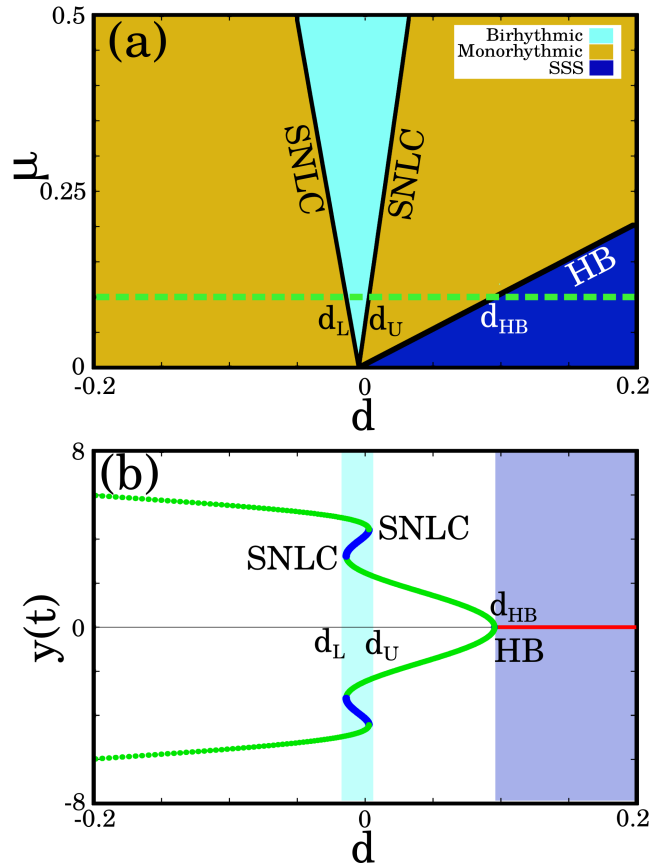}
\caption{(a) Two-parameter bifurcation diagram in $d$-$\mu$ parameter space for $\alpha=0.144$, $\beta=0.005$, (b) one parameter bifurcation diagram with $d$ as control parameter with $\mu=0.1$ [along the broken horizontal line in (a)]. The width of the birhythmic zone is ($d_U-d_L$).}
\label{bif_energy}
\end{figure}

To investigate the detailed bifurcation scenario in the system we use the continuation package XPPAUT\cite{xpp} in the $d-\mu$ parameter space. The two-parameter bifurcation diagram in the $d-\mu$ plane is shown in Fig.~\ref{bif_energy}(a). 
The parameter values are $\alpha=0.144$, $\beta=0.005$, $\Omega_0=1$, $\eta_0=0.1$, $\eta_1=0.25$ and $\gamma=0.2$: this set of parameter values confirm birhythmic oscillations in the uncontrolled system. From Fig.~\ref{bif_energy}(a) we observe that the $d-\mu$ plane is divided by SNLC curves and a HB curve. The birhythmic regime exists between two SNLC curves [cyan (lighter gray) zone]. In this zone three LCs coexist, one with larger amplitude, another with lower amplitude and the third one is unstable. The SNLC curves also govern the transition between the birhythmic zone to the monorhythmic one. The monorhythmic zones are shown by the yellow (light gray) region. The HB line governs the transition between LC and the stable steady state (SSS) [blue (dark) zone]. The HB line exactly matches with the analytically obtained result of Eq.~\eqref{hopfeq}. It may be noted that for $d=0$ (i.e., the uncontrolled case) the system is in a birhythmic regime for all values of $\mu$ for the chosen set of other parameter values.

\begin{figure}
\includegraphics[width=0.45\textwidth]{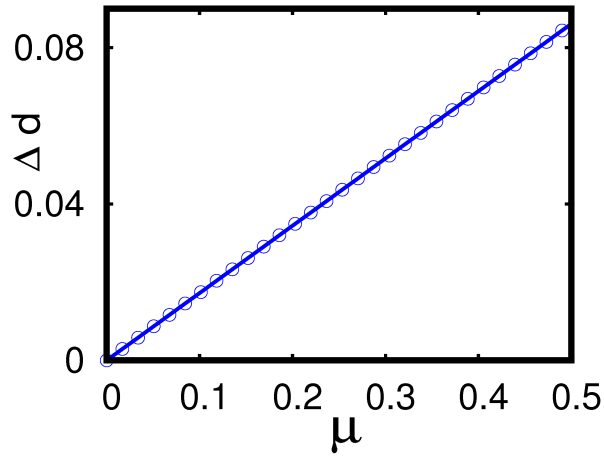}
\caption{The hysteresis width $(\Delta d)$ vs. $\mu$ plot (for other parameters see text).}
\label{hys_mu}
\end{figure}

Next, we take $\mu=0.1$ (horizontal broken line in Fig.~\ref{bif_energy}(a)) and vary $d$. The resulting one-parameter bifurcation diagram is shown in Fig.~\ref{bif_energy}(b). The system is in the birhythmic zone for $d=0$. An increase in $d$ brings the system to  a monorhythmic zone through an SLNC for $d>d_U$. Here the limit cycle is a small-amplitude limit cycle. Further increase in $d$ causes the small amplitude LC to loose stability through an inverse Hopf bifurcation and the system rests in a stable steady state. Therefore, increasing the positive value of the control parameter brings the birhythmic oscillator to a monorhythmic one with small-amplitude oscillation and eventually to a stable steady state. On the other hand, decreasing value of $d$ in negative direction causes the system to experience an SNLC at $d=-d_L$ and monorhythmic oscillation with larger amplitude emerges. To summarize, the proper choice of the control parameter $d$ may cause the system to induce monorhythmicity either with  a small amplitude LC (for $d_U<d<d_{HB}$) or a large amplitude LC (for $d<-d_L$). It is to be noted that, there is a hysteresis around $d=0$ with a width of $\Delta d=d_U-d_L$, which is shown by the cyan (lighter grey) zone in Fig.~\ref{bif_energy}(b). In this zone the state of the system depends on the choice of the initial conditions. It may be noted that the hysteresis zone increases with increasing the value of $\mu$. 
We plot the hysteresis width $\Delta d$ with $\mu$ in Fig.~\ref{hys_mu} for the aforementioned parameters, which shows that the hysteresis width increases with an increase in $\mu$. Therefore, to eliminate the birhythmicity a higher feedback strength ($d$) is necessary for higher values of $\mu$.

\begin{figure}
\includegraphics[width=0.49\textwidth]{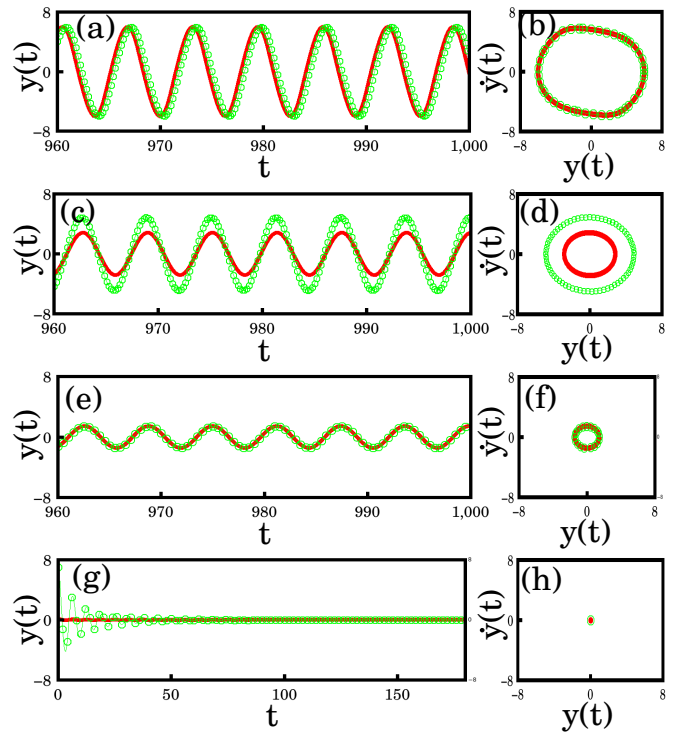}
\caption{Time series and phase plane plots. (a-b) Large-amplitude LC for $d=-0.2$, (c-d) Birhythmic oscillations for $d=-0.01$, (e-f) Small-amplitude oscillation for $d=0.05$ and (g-h) stable steady state for $d=0.2$. The solid line is for initial conditions $(y_0,\dot{y}_0,v_0)=(0.1,0,0.3)$ and the broken line for $(y_0,\dot{y}_0,v_0)=(7,0,0.3)$. Others parameters are: $\mu=0.1$, $\alpha=0.144$, $\beta=0.005$, $\Omega_0=1$, $\eta_0=0.1$, $\eta_1=0.25$, $\gamma=0.2$.}
\label{energy_ts}
\end{figure}

To demonstrate the effectiveness of the control scheme, we plot the time series and the phase plane diagrams for different values of the control parameter $d$ (Fig.~\ref{energy_ts}). Here we use two sets of initial conditions, namely, $\mathcal{I}_1=(0.1,0,0.3)$, which is around the origin, targeting to have the smaller amplitude LC and the other $\mathcal{I}_2=(7,0,0.3)$, away from origin, targeting the larger amplitude LC.
The solid line indicates the result for $\mathcal{I}_1$ and the line with hollow circles that for $\mathcal{I}_2$. Figure \ref{energy_ts}(a) shows time series and Fig.~\ref{energy_ts}(b) shows the phase plane plot for $d=-0.2$ (i.e., $d<-d_L$). We observe that irrespective of the initial conditions the system always shows a large amplitude LC. Figure~\ref{energy_ts}(c-d) demonstrate the scenario for $d=-0.01$, i.e., $-d_L<d<d_U$, for which the system is in the birhythmic region. The occurrence of LCs with two different amplitudes confirms the presence of birhythmicity in the system. For $d>d_U$ there exists only small amplitude LC. This is shown for $d=0.05$ in Fig.~\ref{energy_ts}(e-f). Finally, the oscillation in the system ceases to a stable fixed point for $d>d_{HB}$: This is demonstrated for $d=0.2$ in Fig.~\ref{energy_ts}(g-h).

\begin{figure}
\includegraphics[width=0.49\textwidth]{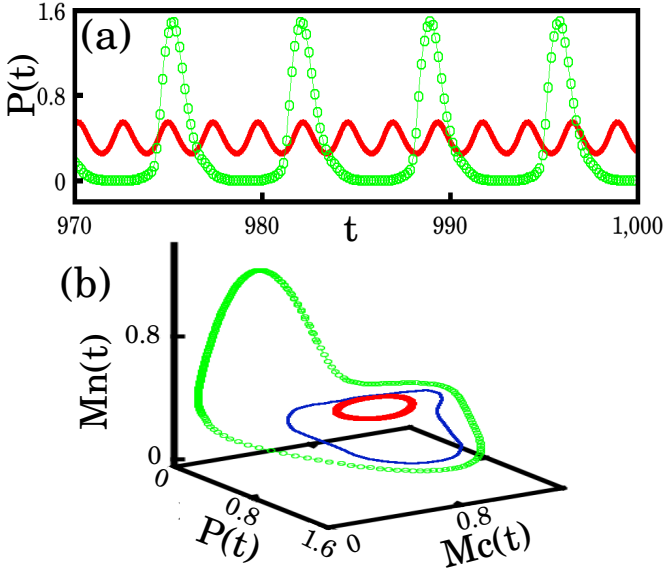}
\caption{(a) Time series, (b) phase plane plot in $P(t)$-$Mc(t)$-$Mn(t)$ space for $d=0$. The small-amplitude LC for IC $\mathcal{I}_1$ and the large-amplitude LC for $\mathcal{I}_2$ (see text). The trajectory in blue in the middle represents  the unstable LC. Other parameters are: $k_p=5$, $K_p=0.2$, $d_p=2.5$, $k_{Mc}=0.1$, $k_{Mc}^\prime=1.2$, $K_{Mc}=0.4$, $k_{in}=0.45$, $k_{in}^\prime=0.4$, $K_{Mn}=0.1$, $d_{Mc}=0.6$, $V_r=10$, $d_{Mn}=1.9$ and $n=6$.}
\label{p53_bi_ts}
\end{figure}
\section{The p53-Mdm2 network: OAK model}
\label{sub:oak}
The control of proliferation of abnormal cells by protein in mammals is modeled through the p53-Mdm2 network, which is called the OAK model originally proposed by by Abou-Jaoud\'{e} et al.\cite{Abou09} The OAK model describes the interaction between p53, cytoplasmic Mdm2 and nuclear Mdm2 \cite{Ciliberto05}. A detailed description of the model may be found in Ref.~\onlinecite{Abou09,p5311}. Nuclear Mdm2 accelerates the degradation of p53 by ubiquitination and by blocking its functional activity. p53 enhances the transcription of gene MDM2 and thus regulates cytoplasmic Mdm2 level. The translocation of Mdm2 from the cytoplasm to the nucleus is inhibited by p53. Although the actual model consists of a 4-dimensional differential equations But in Ref.~\onlinecite{p5311} the model has been reduced to a 3-dimensional one. We apply the self-feedback control term $-d Mc$ and rewrite the equations as follows
\begin{eqnarray}\label{oakeq}
\frac{dP}{dt}&=&k_P\frac{K_P^n}{K_P^n+Mn^n}-d_p P,\nonumber\\
\frac{dMc}{dt}&=&k_{Mc}+k'_{Mc}\frac{P^n}{K_{Mc}^n+P^n}\nonumber\\
&&~~~-\bigg(k_{in}-k'_{in}\frac{P^n}{K_{Mc}^n+P^n}\bigg)Mc\nonumber\\
&&~~~~~~~~~~~~~-d_{Mc}Mc-d Mc,\nonumber\\
\frac{dMn}{dt}&=&V_r\bigg(k_{in}-k'_{in}\frac{P^n}{K_{Mc}^n+P^n}\bigg)Mc\nonumber\\
&&~~~~~~~~~~~~~~~~~~~-d_{Mn}Mn,
\end{eqnarray}
where $P$, $Mc$ and $Mn$ represent the concentrations of p53, cytoplasmic Mdm2 and nuclear Mdm2, respectively. Here the term $d Mc$ in the second equation is the proposed self-feedback. {\color{black} The schematic of the control scheme is represented in Fig.~\ref{oak_mech}.
\begin{figure}
\includegraphics[width=0.45\textwidth]{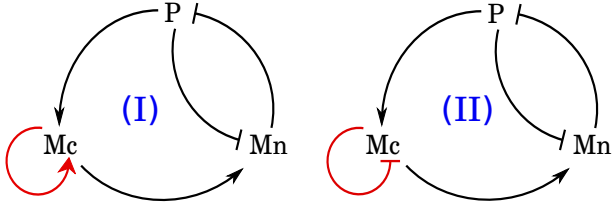}
\caption{Schematic representation of the p53-Mdm2 network with self-feedback. (I) condition for negative $d$ (positive interaction), (II) condition for positive $d$ (negative interaction).}
\label{oak_mech}
\end{figure}
The self-feedback occurs in $Mc$: a portion of $Mc$ is feed to itself (red line). The negative $d$ invokes the positive interaction and is shown by the sharp arrowhead in Fig.~\ref{oak_mech}(I). The positive $d$ represents the negative interaction, i.e., suppression of $Mc$ which is shown in Fig.~\ref{oak_mech}(II).} 

\begin{figure}
\includegraphics[width=0.49\textwidth]{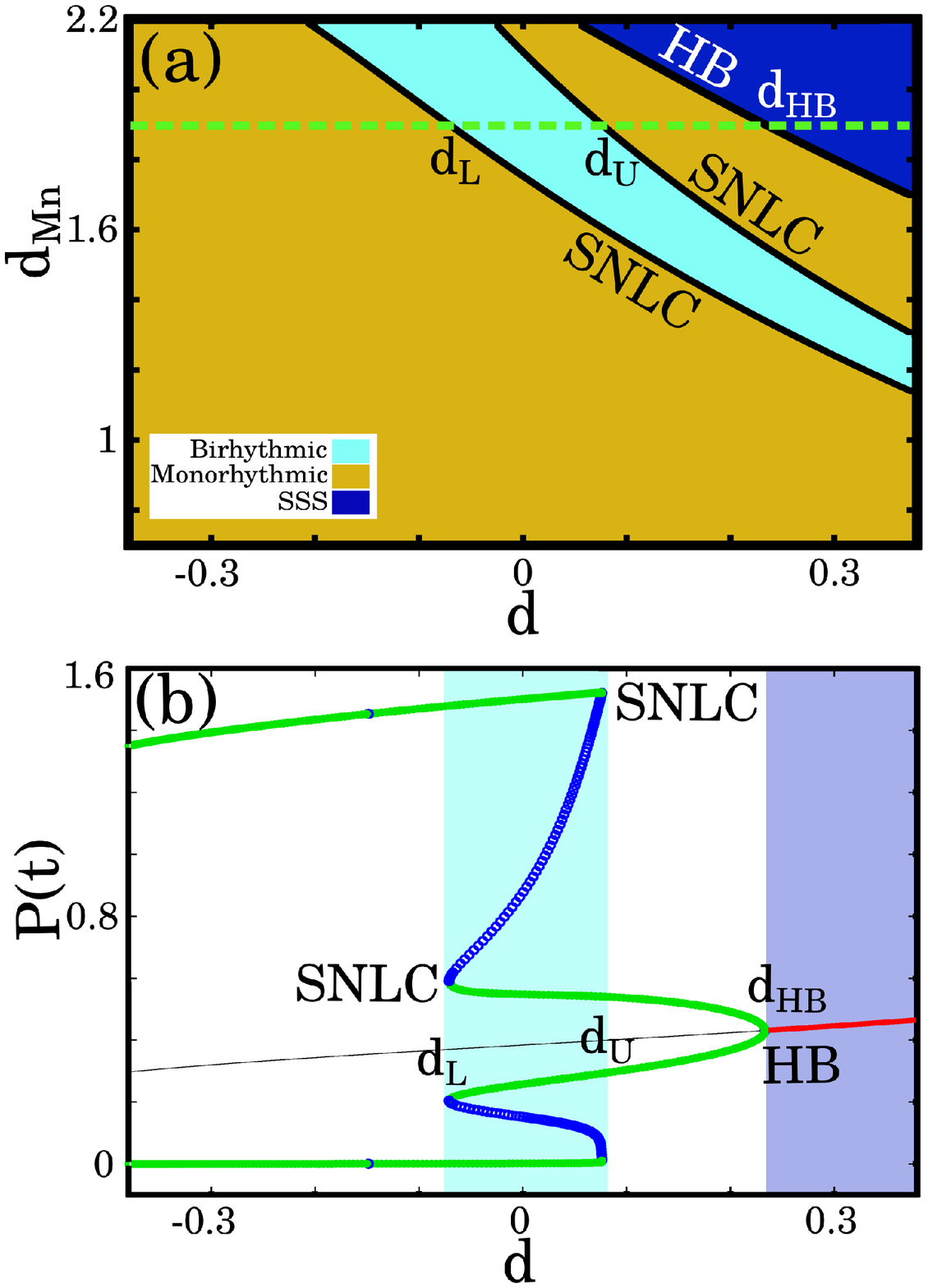}
\caption{(a) Two-parameter bifurcation diagram in $d-d_{Mn}$ parameter space. (b) One parameter bifurcation diagram with $d$ as the control parameter for $d_{Mn}=0.19$ (The broken horizontal line in (a)). Other parameters are same as Fig.~\ref{p53_bi_ts}.}
\label{p53_twopar}
\end{figure}

For $d=0$ we get the original system which shows birhythmicity for the parameter set: $k_p=5$, $K_p=0.2$, $d_p=2.5$, $k_{Mc}=0.1$, $k_{Mc}^\prime=1.2$, $K_{Mc}=0.4$, $k_{in}=0.45$, $k_{in}^\prime=0.4$, $K_{Mn}=0.1$, $d_{Mc}=0.6$, $V_r=10$, $d_{Mn}=1.9$ and $n=6$. The small amplitude LC arising from this system has the initial condition $\mathcal{I}_1\equiv (P(0),Mc(0),Mn(0))\in (0.6,0.3,0.4)$ and the large amplitude LC is a consequence of the initial condition $\mathcal{I}_2\equiv (P(0),Mc(0),Mn(0))\in (3,0.3,0.2)$. The time series and phase plane plots for the original system (i.e., $d=0$ in Eq.\eqref{oakeq}) is shown in Fig.~\ref{p53_bi_ts}(a) and (b), respectively.
To control the birhythmicity, we have applied the self-feedback scheme by using $d\neq 0$. We analyze the bifurcation scenario appearing in this system. The two-parameter bifurcation diagram in the $d-d_{Mn}$ parameter space is shown in Fig.~\ref{p53_twopar}(a). The yellow (light gray) zone shows the monorhythmic zone. The cyan (lighter gray) zone gives the birhythmic zone and the blue (dark) zone is the zone of a stable steady state (SSS).
In this case also we find that the space is divided by SNLC bifurcations and a supercritical Hopf bifurcation. The global SNLC bifurcation distinguishes between the birhythmic and monorhythmic states. The stable limit cycle looses stability through an inverse Hopf bifurcation and a stable steady state emerges.

To understand the scenario in more detail, we draw the one parameter bifurcation diagram sweeping the control parameter $d$ by fixing $d_{Mn}=1.9$ as shown in Fig.~\ref{p53_twopar}(b). The variation of $d$ is considered along the broken yellow horizontal line in Fig.~\ref{p53_twopar}(a). For $d=0$, the system is in the birhythmic zone for the preferred set of parameter values. Increase in $d$ brings the system to monorhythmic oscillations through a global SNLC for $d>d_U$. This monorhythmic oscillation is of small amplitude. Further increase in $d$ causes the stable LC to loose its stability and a stable steady state emerges through a Hopf bifurcation at $d=d_{HB}$. Decrease in $d$ below zero causes the system to enter the monorhythmic zone again but with large amplitude oscillation for $d<-d_L$. From the one-parameter bifurcation diagram we see that there is a hysteresis zone of width $\Delta d=(d_U-d_L)$, which is governed by the SNLC curves.

\begin{figure}
\includegraphics[width=0.49\textwidth]{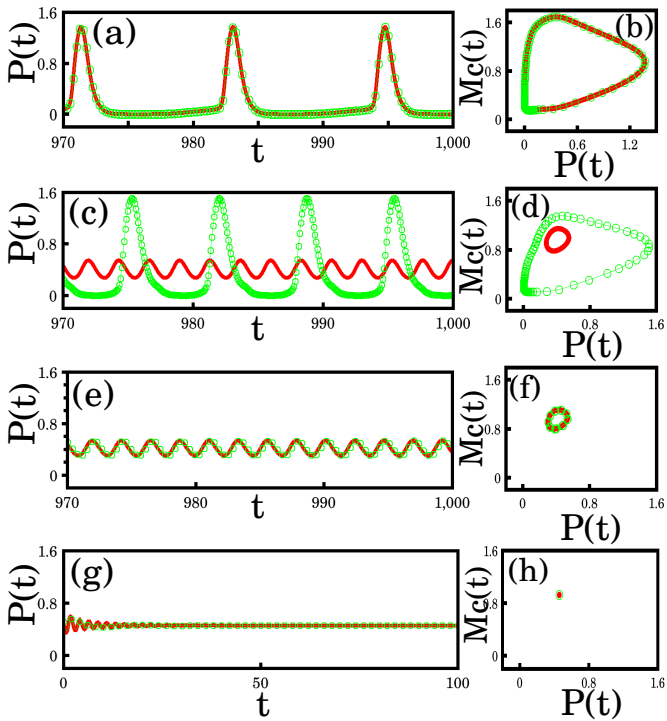}
\caption{Time series and phase plane plots for different values of the control parameter $d$. (a) (time series) and (b) (phase plane) show large amplitude oscillation for $d=-0.35$, (c-d) the birhythmic oscillations for $d=0.04$, (e-f) shows the small-amplitude oscillation for $d=0.1$, and (g-h) the stable steady state for $d=0.35$. The diagram is drawn along the broken horizontal line in Fig.~\ref{p53_twopar}(a). The solid line corresponds to initial conditions: $(P(0),Mc(0),Mn(0)=(0.6,0.3,0.4)$ and the line with hollow circles corresponds to $(P(0),Mc(0),Mn(0)=(3,0.3,0.2)$. Other parameters are same as Fig.~\ref{p53_twopar}.}
\label{p53_ts}
\end{figure}

Finally, we draw the time series and phase plane plots of the system for different values of the control parameter $d$ in Fig.~\ref{p53_ts}. The solid line shows the case of initial condition $\mathcal{I}_1$, and the line with hollow circles shows the same for $\mathcal{I}_2$, respectively.
Fig.~\ref{p53_ts}(a) gives the time series and (b) shows the phase plane plots for $d=-0.35$. Both LCs are of the same large amplitude irrespective of initial conditions indicating monorhythmic oscillation. For $d=0.04$, the system is in the bithythmic zone shown in Fig.~\ref{p53_ts}(c-d). Increasing $d$ brings the system to the monorhythmic zone with small amplitude oscillation for $d>d_U$. The situation for $d=0.1$ is shown in Fig.~\ref{p53_ts}(e-f). Finally the system enters the stable steady state for $d>D_{HB}$: Fig.~\ref{p53_ts}(g-h) demonstrates the scenario for $d=0.35$. Thus, it is worth noting that the birhythmicity may be controlled and eliminated with proper choice of the control parameter $d$. 

\section{The Glycolysis model: modified Decroly-Goldbeter model}
\label{sub:glyco}
\begin{figure}
\centering
\includegraphics[width=0.4\textwidth]{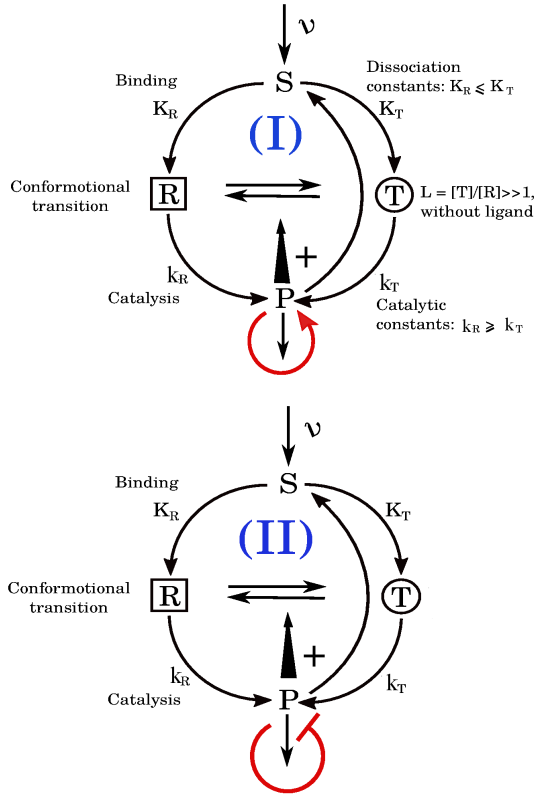}
\caption{Schematic representation of the glycolysis model with self-feedback. (I) condition for negative $d$ (positive interaction), (II) condition for positive $d$ (negative interaction). See Ref.~\onlinecite{Goldbeter96} (chapter 2) and Ref.~\onlinecite{Dsray04} for a detailed description of the pathways and the relevant thermodynamic and kinetic parameters.}
\label{glyco_mech}
\end{figure}

Enzymatic oscillations with periodicity (of several minutes) have two-fold interest in biology, first, in metabolic pathways and second, as general models for biological rhythms. These examples include glycolytic oscillations in yeast and muscle and the periodic synthesis of cAMP during the aggregation of the slime mold {\em Dictyostelium discoideum} \cite{Martiel87, Gorbunova02}. \citet{Dsray04} consider a product-activated enzyme model which is a modified version of the well known Decroly-Goldbeter model\cite{Goldbeter96}. According to them, the allosteric enzymes consist of multiple identical subunits. These subunits undergo conformational transition between more reactive (R) and less reactive (T) states. Here the substrate ($(S)$) injection rate $\nu$ is constant. The product $P$ is resulted from the bindings of $S$ with $R$ and $T$ states of the enzyme. Then the product is removed with a rate proportional to its concentration, which results in a positive feedback and activates the $T$ to $R$ transitions. When the product $P$ gives a positive feedback to substrate $S$ birhythmicity results. The system dynamics is described by the following equations along with the self-feedback mechanism
\begin{eqnarray}
\frac{d\alpha}{dt}&=& \nu-\sigma\phi(\alpha,\gamma)+\frac{\sigma_i \gamma^n}{K^n+\gamma^n},\nonumber\\
\frac{d\gamma}{dt}&=& q\sigma\phi(\alpha,\gamma)-K_s\gamma-\frac{q\sigma_i \gamma^n}{K^n+\gamma^n}-d \gamma.\label{goldeq}
\end{eqnarray}
with,
\begin{equation}\label{goldfn}
\phi(\alpha,\gamma)=\frac{\alpha(1+\alpha)(1+\gamma)^2}{L+(1+\alpha)^2(1+\gamma)^2}.
\end{equation}
Here $\alpha$ is the normalized substrate concentration and $\gamma$ is the normalized product concentration. The self-feedback control is provided by the $-d\gamma$ term in the second subequation of \eqref{goldeq}. {\color{black} The schematic of the pathways is shown in Fig.~\ref{glyco_mech} \cite{Goldbeter96,Dsray04}; the self-feedback term is shown using red lines in Fig.~\ref{glyco_mech}. According to Ref.~\onlinecite{Goldbeter96} the reaction product ($P$) leaves the system at a rate proportional to its
concentration. Therefore, in our case a positive self-feedback represented by Fig.~\ref{glyco_mech} (I) means an accumulation of the  reaction product ($P$), whereas, a negative feedback represented by Fig.~\ref{glyco_mech} (II) means an extraction of the reaction product ($P$). Here we make use of the kinetic assumption that for a step catalysed by a Michaelian enzyme is not saturated by its substrate \cite{Goldbeter96}. Earlier, theoretical \cite{Goldbeter96,Dsray04} and experimental observations \cite{cell} of the glycolytic oscillations were made by changing the substrate injection rate $\nu$. However, in the present case we keep $\nu$ constant and study the effect of the control on the reaction product ($P$).}  

\begin{figure}
\includegraphics[width=0.49\textwidth]{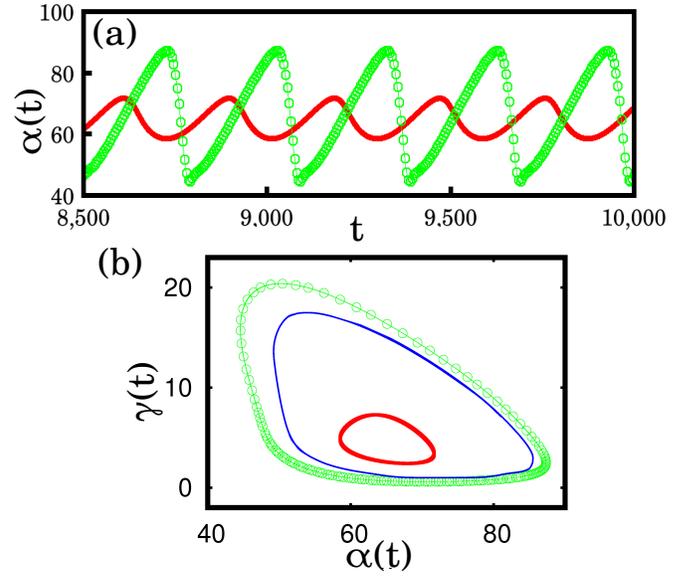}
\caption{(a) Time series, (b) phase plane plot in $\alpha(t)$-$\gamma(t)$ space for $d=0$. The small-amplitude LC for IC $\mathcal{I}_1$ and the large-amplitude LC for $\mathcal{I}_2$ (see text). The trajectory in blue in the middle represents unstable LC. Other parameters are: $\nu=0.255$, $q=1.0$, $K_s=0.06$, $L=3.6\times 10^6$, $\sigma=10$, $\sigma_i=1.3$, $n=4$ and $K=10.0$.}
\label{goldbeter_bi_ts}
\end{figure}

For $d=0$ the system is reduced to the original one and is capable of showing birhythmicity for the following set of parameters: $\nu=0.255$, $q=1.0$, $K_s=0.06$, $L=3.6\times 10^6$, $\sigma=10$, $\sigma_i=1.3$, $n=4$ and $K=10.0$. The system exhibits large amplitude LC for the initial conditions $\mathcal{I}_1\equiv (\alpha(0),\gamma(0))\in(100,5)$ and small amplitude LC for $\mathcal{I}_2\equiv (\alpha(0),\gamma(0))\in(80,5)$, respectively. The time series and phase plane plots for the original system (i.e., $d=0$ in Eq.\eqref{goldeq}) are shown in Fig.~\ref{goldbeter_bi_ts}(a) and (b), respectively.

\begin{figure}
\includegraphics[width=0.49\textwidth]{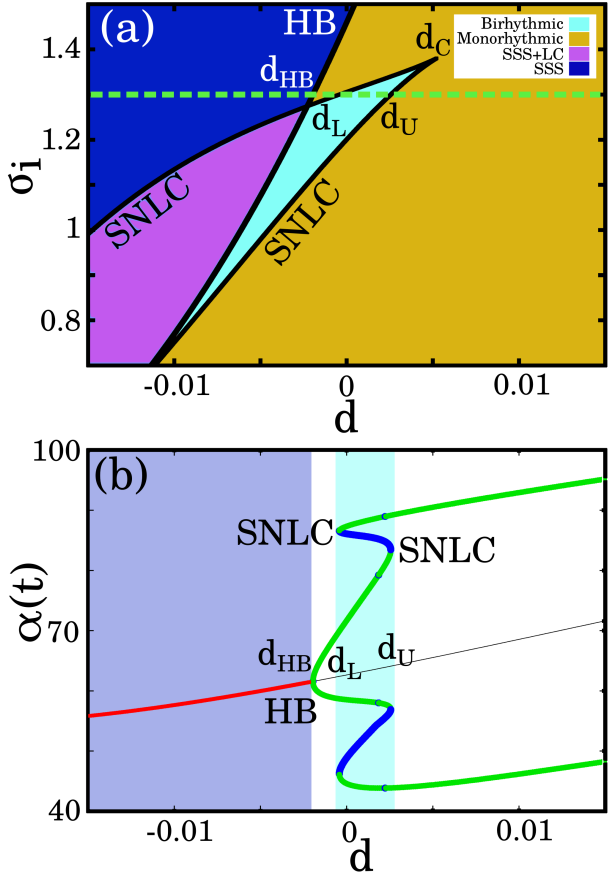}
\caption{(a) Two-parameter bifurcation diagram in $d-\sigma_i$ parameter space. (b) One parameter bifurcation diagram with $d$ as the control parameter for $\sigma_i=1.3$ (The broken horizontal line in (a)). Other parameters are same as Fig.~\ref{goldbeter_bi_ts}.}
\label{goldbeter_bif}
\end{figure}

The control is active for $d\neq 0$. We investigate the possible bifurcation scenario appearing in the system. In Fig.~\ref{goldbeter_bif}(a) we present the two-parameter bifurcation diagram in the $d-\sigma_i$ parameter space. It is divided in birhythmic, monorhythmic and stable steady state zones by  SNLC bifurcation and Hopf bifurcation curves. The SNLC bifurcation governs the transition between birhythmicity and monorhythmicity and the (inverse) Hopf bifurcation brings the system to a stable steady state. The yellow (light gray) zone in Fig.~\ref{goldbeter_bif}(a) shows the monorhythmic oscillatory zone. The cyan (lighter gray) zone shows the birhythmic regime and the blue (dark) zone the stable steady state. Here another interesting bistable state exists, namely, the coexistence of one large amplitude LC and one stable steady state between the Hopf and SNLC curves (the purple zone).

To have a better understanding of this, we further draw the one-parameter bifurcation diagram with $d$ as the parameter (sweeping $d$ along the broken horizontal line in Fig.~\ref{goldbeter_bif}(a)). The diagram is shown in Fig.~\ref{goldbeter_bif}(b). The birhythmic zone is presented by the shaded region in the figure. 
For $d<d_{\mbox{HB}}$ the system is in stable steady states. At $d=d_{\mbox{HB}}$ a supercritical Hopf bifurcation occurs and oscillation of small amplitude emerges. For $d_{\mbox{HB}}<d<d_L$ the system is in a monorhythmic zone with small amplitude LC. The birhythmic zone lies between $d_L<d<d_U$. For $d>d_U$ the system shows monorhythmicity with large amplitude oscillation. Note that the variation of $d$ causes the system to undergo a transition, which is inverse to the previous two models discussed in earlier subsections. It is interesting to note that, there is a hysteresis governed by the SNLC bifurcations (cyan zone) with the hysteresis width $\Delta d=(d_U-d_L)$. The SNLC curves meet each other at $d=d_C$ indicating the presence of a codimension-2 cusp type of bifurcation.

\begin{figure}
\includegraphics[width=0.49\textwidth]{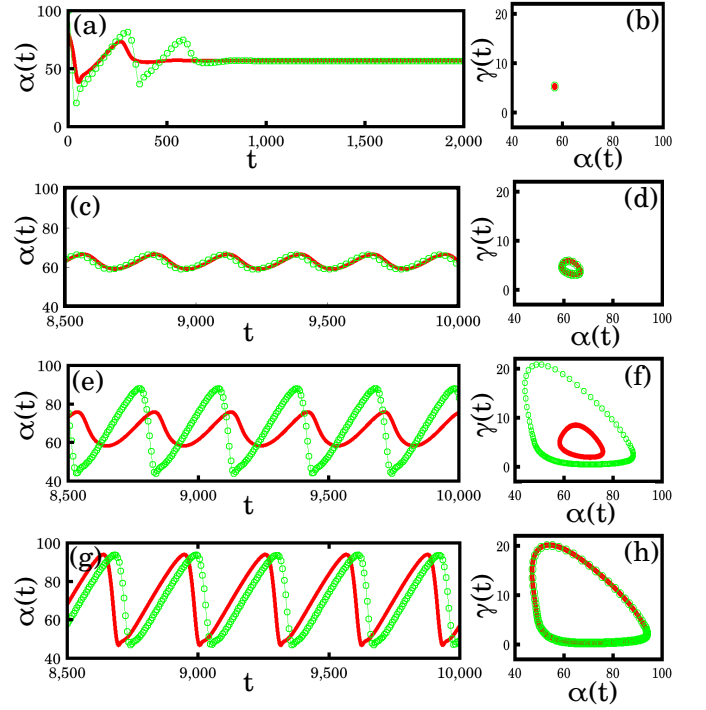}
\caption{Time series and phase plane plots for different values of the control parameter $d$. (a) (time series) and (b) (phase plane) show stable steady state for $d=-0.012$, (c-d) the small amplitude oscillation for $d=-0.001$, (e-f) the birhythmic oscillation for $d=0.001$, and (g-h) the large amplitude oscillation for $d=0.012$. The diagram is drawn along the broken horizontal line in Fig.~\ref{p53_twopar}(a). The solid line corresponds to initial conditions: $\alpha(0)=100$, $\gamma(0)=5$ and the line with hollow circles corresponds to $\alpha(0)=80$, $\beta(0)=5$. Other parameters are same as Fig.~\ref{goldbeter_bi_ts}.}
\label{goldbeter_ts}
\end{figure}

Finally, we demonstrate the time series of the system for different values of the control parameter $d$ with two different initial conditions, namely, $\mathcal{I}_1$ and $\mathcal{I}_2$ in Fig.~\ref{goldbeter_ts}. The solid line in the figure indicates the result for $\mathcal{I}_1$ and the line with hollow circles that for $\mathcal{I}_2$. The two initial conditions are chosen in such a way that the system shows small amplitude oscillation for $\mathcal{I}_1$ and large amplitude oscillation for $\mathcal{I}_2$ when $d=0$ (cf. Fig.~\ref{goldbeter_bi_ts}).
Fig.~\ref{goldbeter_ts}(a,b) shows time series and phase plane plots, respectively, for $d=-0.012$ $(d<d_{HB})$. The system rests in a stable steady point. A supercritical Hopf bifurcation occurs at $d=d_{HB}$ and the system revives from its dormant state and LC of small amplitude emerges. The situation for $d=-0.001$ is shown in Fig.~\ref{goldbeter_ts}(c-d). In the range $d_L<d<d_U$ the system exhibits a birhythmic nature. Fig.~\ref{goldbeter_ts}(e-f) shows the case for $d=0.001$. Finally, the system creates a large amplitude LC for $d>d_U$. The scenario for $d=0.012$ is shown in Fig.~\ref{goldbeter_ts}(g-h). Thus we can conclude that, like the two previous models, the control scheme can effectively eliminate the birhythmic behavior and may lead to a preferred monorhythmic state.

\section{Conclusions}
\label{sec:conc}
In conclusion, we have {\color{black} studied the effect of} self-feedback mechanism on birhythmic oscillations. {\color{black} We have shown that the self-feedback through a proper variable} is able to eliminate birhythmic oscillations and select monorhythmic oscillation of either large amplitude or small amplitude, as desired. We have successfully applied the self-feedback mechanism to three realistic models from different branches of natural science, e.g., an energy harvesting system and biochemical models such as the OAK model and glycolysis model. A rigorous analysis using harmonic decomposition and energy balance method have established the efficacy of the self-feedback mechanism. Further, we have explored the possible bifurcation scenarios to get a deep understanding of the genesis of monorhythmic oscillation that comes out of birhythmic oscillation. In comparison with our previously proposed control scheme in Ref.~\onlinecite{Biswas2016control} the proposed self-feedback control technique uses self-feedback of only a single variable, hence the physical implementation of this control scheme is comparatively easy. We believe that our proposed coupling scheme is general enough to be applied effectively to control birhythmicity in several physical and biochemical processes.  

\begin{acknowledgments}
{\color{black} Authors thankfully acknowledge the insightful suggestions by the anonymous referees.} DB acknowledges CSIR, New Delhi, India. TB acknowledges Science and Engineering Research Board (Department of Science and Technology, India) [grant No. SB/FTP/PS-05/2013]. DB acknowledges Haradhan Kundu, Department of Mathematics, University of Burdwan, for his useful suggestions regarding computations. 
\end{acknowledgments}

{\color{black}
\appendix
\section{Stability analysis using Poincar\'e-Lindstedt method}
\label{app:plm}
As we discussed in the main text, the harmonic decomposition technique is not an asymptotic method and in this context the Poincar\'e-Lindstedt technique is a much more suitable technique for weakly nonlinear systems. Since we are interested to find out the {\it amplitude equation} of Eq.~\eqref{energycoupsingle}, therefore in the following we show that both the techniques give similar {\it amplitude equations}.

Let us consider the system equation Eq.~\eqref{energycoupsingle}. With $\Omega_0=1$ it becomes
\begin{equation}\label{app_energycoupsingle}
\ddot{y}+\mu F(\dot{y})+y+d\dot{y}=0.
\end{equation}
With $F(\dot{y})$ given by Eq.~\eqref{energyfn}. We consider the solution to be of the form
\begin{equation}\label{appy}
y(\tau)=y_0(\tau)+\mu y_1(\tau)+\mu^2 y_2(\tau)+\cdots,
\end{equation}
where $\tau\equiv\omega t$ and $\omega$ is a known frequency and $y_i(\tau)$, ($i=1,2,\dots$) is the periodic function of periodic $2\pi$. We also write
\begin{equation}\label{appom}
\omega=1+\mu\omega_1+\mu^2\omega_2+\cdots,
\end{equation}
where $\omega_i$ are unknown ad needed to determine. Further, we decompose the control parameter $d$ as
\begin{equation}\label{appd}
d=\mu d_1+\mu^2 d_2+\cdots
\end{equation}
Substituting Eqs.~\eqref{appy}, \eqref{appom} and \eqref{appd} into Eq.~\eqref{app_energycoupsingle} and equating the coefficients of different powers of $\mu$, we get

Coeff. of $\mu^0$:
\begin{equation}\label{appm0}
\ddot{y}_0+y_0=0.
\end{equation}

Coeff. of $\mu^1$:
\begin{equation}\label{appm1}
\ddot{y}_1+y_1=-2\omega_1\ddot{y}_0-F(\dot{y}_0)-d_1 y_0.
\end{equation}

Coeff. of $\mu^2$:
\begin{equation}\label{appm2}
\begin{split}
\ddot{y}_2&+y_2=\\
& -2\omega_1\ddot{y}_1-d_1 y_1-(\omega_1^2+2\omega_2)\ddot{y}_0-\omega_1\dot{y}_0^3\\
&-\beta\omega_1\dot{y}_0^7-\dot{y}_0^2\dot{y}_1-\beta\dot{y}_0^6\dot{y}_1\\
&+(1+\alpha \dot{y}_0^4)(\omega_1 \dot{y}_0+\dot{y}_1)-d_2 y_0.
\end{split}
\end{equation}
The solution of Eq.~\eqref{appm0} is given by
\begin{equation}\label{appy0}
y_0(\tau)=Ae^{i\tau}+c,
\end{equation}
where $A$ is the complex amplitude to be determined. Substitution of Eq.~\eqref{appy0} in Eq.~\eqref{appm2} invokes the solvability condition as 
\begin{equation}\label{appsolv}
2\omega_1+i(1-A\bar{A}+2\alpha A^2\bar{A}^2-5\beta A^3\bar{A}^3-d_1)=0.
\end{equation}

Equating the real and imaginary parts of the above equation
\begin{equation}\label{appom1}
\omega_1=0,
\end{equation}
and 
\begin{equation}\label{appamp}
1-|A|+2\alpha |A|^4-5\beta |A|^6-d_1=0.
\end{equation}
Therefore, with $A$ replaced by $\frac{A}{2}$, the {\it amplitude equation} of Eq.~\eqref{appamp} is similar to that obtained in Eq.~\eqref{ampeq1} (using the harmonic decomposition technique). Apart from amplitude equation one can show $\omega_2\neq 0$ (note that $\omega_1=0$) by applying the similar steps carried above but for the coefficient of $\mu^2$: However, this analysis does not affect our amplitude equations.}

\providecommand{\noopsort}[1]{}\providecommand{\singleletter}[1]{#1}%
\end{document}